\documentstyle[prl,aps]{revtex} 
\begin{document} 
\draft
\title{Interlayer Josephson vortices in the high-$T_c$\\
superconducting cuprates}
\author{Behnam Farid\thanks{Max-Planck-Institut f\"ur Festk\"orperforschung,
Heisenbergstra\ss e 1,
70569 Stuttgart, Federal Republic of Germany.
{Electronic address: farid@audrey.mpi-stuttgart.mpg.de} } }
\maketitle
\widetext

\vskip 30pt
\widetext
{\Huge\bf A}ccording to the inter-layer tunneling (ILT) theory of
superconductivity for the doped high-transition-temperature (high-$T_c$) 
cuprates by Anderson and co-workers (1), the superconducting 
condensation energy is {\sl almost entirely} equal to the Josephson 
coupling energy $E_J^0$. On the basis of this hypothesis, for single-layer 
cuprates Anderson (2) has derived (in the CGS-Gaussian units) $E_J^0 
\approx \Phi_0^2/[16\pi^3 d^2 \lambda_c^2]$, where $\Phi_0$ stands for 
the superconductor quantum of magnetic flux, $d$ the interlayer distance 
and $\lambda_c$ the penetration depth along the $c$-axis, i.e. that normal 
to the CuO planes. For given values for $E_J^0$ and $d$, the latter 
relation provides an estimate for $\lambda_c$.

Within the `elliptical approximation' (EA) of a model for an infinite 
stack of equi-distant and identical layers of superconductors, originally 
introduced by Lawrence and Doniach (3), and subsequently modified by 
Clem (4) and Clem and Coffey (5) [the `LDC model'], $\lambda_c$ is seen 
to determine the size of an isolated inter-layer Josephson vortex, assumed 
to be ellipse-like, parallel to CuO planes. From their measurements of the 
{\sl total} magnetic flux $\Phi_s$ through an $8.2\times 8.2$ $\mu$m$^2$ 
rectangular pick-up loop, corresponding to some isolated inter-layer 
Josephson vortices in Tl$_2$Ba$_2$CuO$_{6+\delta}$ (Tl-2201, which is a 
single-layer compound with $d=11.6$~\AA), combined with a theoretical 
expression concerning the magnetic flux {\sl density} due to an isolated 
Josephson vortex as put forward by Clem and Coffey (5) within the 
above-mentioned EA, Moler, {\sl et al.}, (6) have obtained $\lambda_c 
\approx 22$ $\mu$m, to be compared with $\lambda_c^{\rm ILT} \approx$ 
1 -- 2 $\mu$m (2). This implies that contrary to the basic hypothesis of 
the ILT theory, the Josephson coupling energy in Tl-2201 should merely 
amount to $\approx 0.1$\% of the condensation energy.

We have studied the EA of the LDC model in considerable detail (7). Our 
main observation is that {\sl the LDC model within the EA has {\bf no} 
physical solution}. Consequently the Clem-Coffey solution should be 
spurious. Indeed, our fittings of $\Phi_s/\Phi_0$ based on this solution 
to the experimental data by Moler, {\sl et al.}, (6) beyond the {\sl 
two}-parameter fitting adopted by these authors (with $\lambda_c$ and 
$z_0$ the fitting parameters --- $z_0$ is the distance of the pick-up 
loop from the surface of Tl-2201 parallel to the $ac$-plane), give results 
which vastly deviate from the known physical parameters pertaining to 
Tl-2201. For instance, by considering $\lambda_a$ (the penetration depth 
along the $a$-axis) as an additional fitting parameter, our fitting of 
$\Phi_s/\Phi_0$, based on the Clem-Coffey solution, to experimental data 
as presented in Fig.~2(I) of Ref.~6, yields: $\lambda_c = 16.79$ $\mu$m, 
$\lambda_a = 6.76\times 10^{-4}$ $\mu$m and $z_0 = 1.33$ $\mu$m; the 
standard deviation $\sigma$ of our fitted curve from the experimental 
curve for $\Phi_s/\Phi_0$ amounts to $\sigma = 3.2\times 10^{-3}$. It is 
important to point out that by using $\lambda_c$ and $z_0$ as the sole
fitting parameters, we obtain $\lambda_c = 22.43$ $\mu$m and $z_0 = 2.97$ 
$\mu$m (with $\sigma = 4.3\times 10^{-3}$), in conformity with the results 
reported in Ref.~6, namely $\lambda_c = 22_{-4}^{+6}$ $\mu$m and $z_0 = 
3.0 \pm 0.6$ $\mu$m. The above-presented fitted value for $\lambda_a$ is 
by three orders of magnitude smaller than the accepted value $\lambda_a = 
0.17$ $\mu$m, employed in the $(\lambda_c,z_0)$-fittings. By including 
additionally the interlayer distance $d$ in our fittings, we have obtained 
values for $d$ which are by 15 to 20 times {\sl larger} than the 
experimentally-known value of $d = 11.6$~\AA. We therefore conclude that 
the experimental results by Moler, {\sl et al.}, (6) should be reanalyzed 
within an improved theoretical framework. Not before this, can one consider 
$\lambda_c \approx 22$ $\mu$m as the definitive value for $\lambda_c$ in 
Tl-2201. We should like to emphasize that our present Comment does not 
consider the question whether the above-presented expression for $E_J^0$ 
does justice to the ILT theory or not (in our view {\sl it does 
not}). Consequently, our reservation with regard to assigning the value 
$\approx 22$ $\mu$m to $\lambda_c$ for Tl-2201 may also be considered in
disregard to the consequence of $\lambda_c \approx 22$ $\mu$m for the 
ILT theory.

\vfill
\pagebreak
\vskip 30pt
\centerline{\bf\sc REFERENCES AND NOTES}

\end{document}